\documentclass[letterpaper, 10 pt, conference]{ieeeconf}  

\IEEEoverridecommandlockouts
\usepackage[utf8]{inputenc}
\usepackage{authblk}
\usepackage{graphicx}
\usepackage{fixltx2e}
\usepackage{booktabs}
\usepackage{array}
\graphicspath{ {./Pix/} }

\title{\LARGE \bf
DRDr II: Detecting the Severity Level of Diabetic Retinopathy Using Mask RCNN and Transfer Learning
}

\author[1]{Farzan Shenavarmasouleh}
\author[1]{Farid Ghareh Mohammadi}
\author[2]{M. Hadi Amini}
\author[1]{Hamid R. Arabnia}
\affil[1]{Department of Computer Science, University of Georgia, Athens, GA, USA \authorcr {\{fs04199, farid.ghm, hra\}@uga.edu}\vspace{1.5ex}}
\affil[2]{School of Computing \& Information Sciences, solid lab, Florida International University, Miami, FL, USA \authorcr {\{moamini\}@fiu.edu}\vspace{1.5ex}}

\begin{document}
\maketitle
\thispagestyle{empty}
\pagestyle{empty}

\begin{abstract}
DRDr II is a hybrid of machine learning and deep learning worlds. It builds on the successes of its antecedent, namely, DRDr, that was trained to detect, locate, and create segmentation masks for two types of lesions (exudates and microaneurysms) that can be found in the eyes of the Diabetic Retinopathy (DR) patients; and uses the entire model as a solid feature extractor in the core of its pipeline to detect the severity level of the DR cases. We employ a big dataset with over 35 thousand fundus images collected from around the globe and after 2 phases of preprocessing alongside feature extraction, we succeed in predicting the correct severity levels with over 92\% accuracy.

\vspace{\baselineskip}

\end{abstract}
\section{\textbf{INTRODUCTION}}
\textbf{Motivation:} Diabetic is among the most infamous diseases in the world and it affects millions of people every year. It can cause problems in several organs of the body, one of which is the eyes. When it does, it can cause vision impairment and if not treated professionally it can eventually lead to vision loss. Diabetic Retinopathy (DR), is the name that the experts have given to this subsection of the disease. Patients need to undergo a simple screening to get their fundus images taken by the retinal photography devices in clinics. Experienced clinicians will then need to analyze the pictures and decide on the presence and the significance level of the case by carefully examining countless features in them. Those features happen to be extremely small in many cases (e.g. less than 5 pixels) and can be very difficult and time-consuming to find even for the best-trained eyes. But, the good news is once detected soon enough, laser surgeries can be used as the treatment. Hence, the sooner DR is diagnosed, the more chance the patient has to recover from it and get the most out of his/her eyes.

As the number of diabetic patients is growing annually and each patient needs frequent screenings, more and more images need to be analyzed by experts every day and since us, humans, are prone to mistakes in repetitive tasks, it increases the risk of not diagnosing DR in time that can be crucial to patients health. Besides, research has shown that out of all the screenings conducted annually, only 25.2\% show trays of DR in them and the patient has to be referred to ophthalmologists \cite{massin2008ophdiat}.

Looking at this from a wider perspective, we can see that the whole process is being taken care of sub-optimally on both sides. Doctors spend nearly three-fourths of their time analyzing healthy patients and patients with mild cases of DR are always at the risk of being left undiagnosed. Fortunately, machine learning and deep learning can come to the rescue and help in many areas with their speed and accuracy. Biomedical imaging is not an exception and has always been an area of interest for researchers in these fields.

As mentioned earlier, doctors need to analyze fundus images and decide on the presence and the significance level of DR. In our preceding paper \cite{shenavarmasouleh2020drdr}, we proposed DRDr, a deep learning model derived from a complex Convolutional Neural Network, namely Mask RCNN, that could take a fundus image as the input, and output all the instances of microaneurysms and exudates - two types of lesions that could appear in the eyes of DR patients - present in it, their position in addition to their exact shapes as separate black and white binary masks, and a confidence score for each of them, all in near real-time. Hence, taking care of the former problem and helping doctors in coming to a conclusion about the presence of the DR.

\begin{figure*}[ht]
    \centering
    \includegraphics[width=17.5cm]{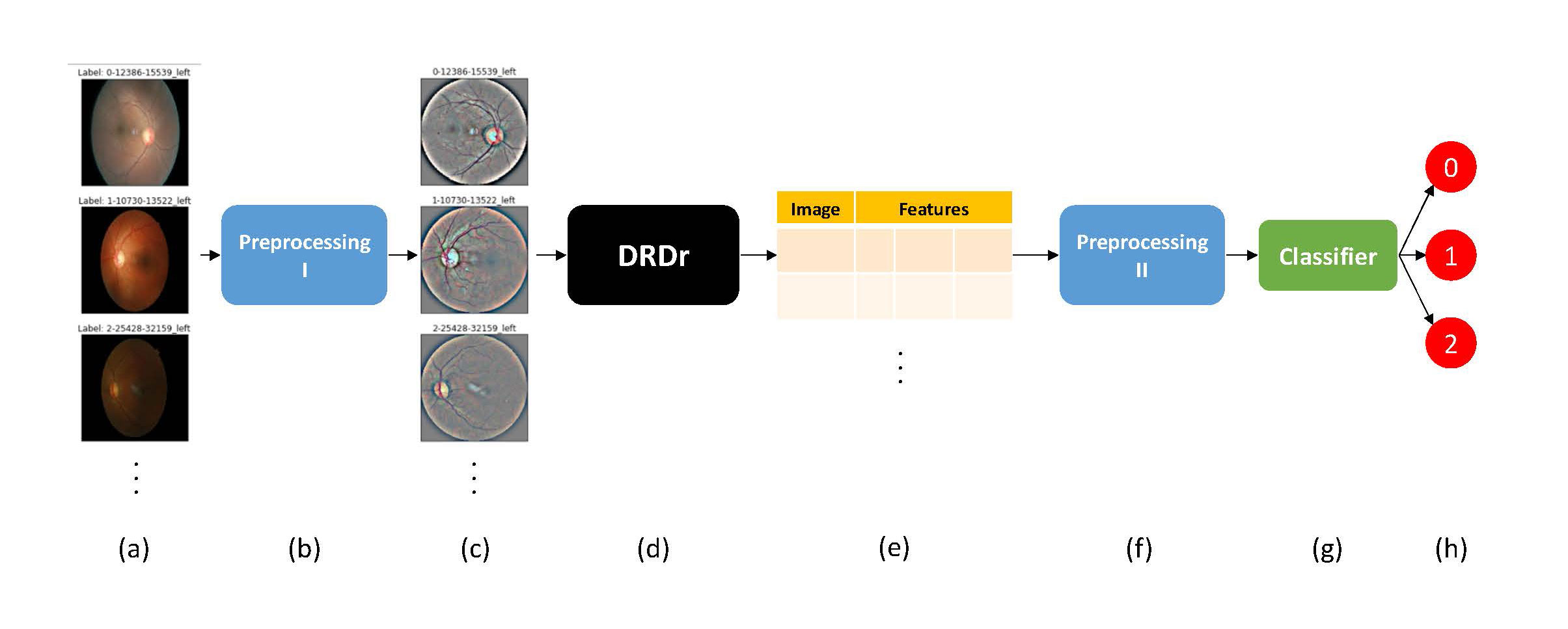}
    \caption{(a) original images in the dataset. (b) perform all the Preprocessing I steps explained in section 3.2. (c) preprocessed images. (d,e) pump all the preprocessed images into the DRDr model and use the output generated by it for all images to create a data frame for features. The steps are explained in section 3.3. (f) apply the Preprocessing II steps explained in section 3.4. (g,h) feed the data frame to an arbitrary classifier and get the severity levels as explained in section 4.
    }
    \label{fig:Fig1}
\end{figure*}

\textbf{Contribution:} In this paper, we propose DRDr II, a hybrid of machine learning and deep learning approaches that uses DRDr as a feature extractor in the core of its pipeline, and with that, it becomes able to tackle the latter problem and classify patients into three severity groups. An overview of the system can be found in Figure \ref{fig:Fig1}.

\textbf{Organization:} The remainder of the paper is organized as follows: First, we start with the related works. Then, we fully explain our several preprocessing steps and feature extraction phase. Next, we show the experiments that we conducted and illustrate the results. Finally, we conclude with the discussion and future work.

\section{\textbf{Related Work}}
An extensive amount of research has been done to find the severity of diabetic retinopathy in patients and classify images into several categories. The majority of them use pure image processing and computer graphics techniques such as Adaptive Histogram Equalization, Discrete Wavelet Transform, Matched Filter Response, Fuzzy C-Means Segmentation, Region Growing, Thresholding \cite{liu1997automatic, ege2000screening, sinthanayothin1999automated} along with other Morphological Processings as feature extractors and pass the result to machine learning classifiers. More than often, authors have decided to go with an ensemble of classifiers in order to make their model more robust and get a better result. Meta-learning has also shown promising results and researchers could benefit from its simplicity and speed as it tries to learn using the least amount of features \cite{mohammadi2020introduction, mohammadi2019promises}.

Traditionally, the problem definition revolved around binary classification, and models were trained to label patients as healthy vs unhealthy. To that end, some merged image features with the contextual data that were available for the patients \cite{decenciere2013teleophta} and a few used Decision Trees, Naïve Bayes, Random Forests, Adaboost, Probabilistic Neural Networks (PNN), K Nearest Neighbors (KNN), and Support Vector Machines (SVM) as ensembles to tackle this issue \cite{bhatia2016diagnosis} \cite{antal2014ensemble} \cite{priya2013diagnosis} \cite{sopharak2010machine}.

Gardner et al. \cite{gardner1996automatic} were one of the firsts who used feed-forward neural networks as the classifier. They converted the fundus images into grayscale and created small tiles out of the entire picture for each of them. The 20x20 tiles were then converted to the vectors and were fed into the neural network to find the deficiency class. Usher et al. \cite{usher2004automated} also employed neural networks and fed candidate lesions, their position, and their type as the inputs of the Neural Network. DRDr II is in line with this work in terms of the employed features.

Releasing of better datasets that provided more severity classes was good news for researchers as they started to level up their models. Same as before, traditional machine learning models were the default option here and SVM, KNN, Gaussian Mixture Model (GMM), and Adaboost were used to classify fundus images into three \cite{lachure2015diabetic}, four \cite{roychowdhury2013dream}, and five \cite{acharya2008application, adarsh2013multiclass} groups. But, as of the past few years, with the rise of deep learning models, researchers could get promising results solely by using image features extracted via Convolutional Neural Networks (CNN) and group patients into two \cite{gargeya2017automated, gulshan2016development}, and five \cite{pratt2016convolutional} classes. 

Inevitably, pure CNN models have the advantage of taking an image as a whole and use it completely as their available features. They try to figure out which parts of the image are best to use and what to look for in images for the sole purpose of getting a better accuracy in the classification task (severity level). So, in theory, the model can converge to a point that it can cognitively understand all the types of lesions and deficiencies, such as cotton wools spots, abnormal growth of blood vessels, hemorrhages, and much more in addition to the two (exudates and microaneurysms) that we were bound to use due to lack of proper databases. With all that said, we will show that with proper techniques, two is all it takes to yield an acceptable 92.55\% classification accuracy.

One big drawback of all the aforementioned related works is that they each used different datasets for their work. Some were public datasets but some were private ones created only for their labs. Also, the size of the datasets varied by a big margin, as they could contain as few images as 39 in total (extremely prone to overfitting), all the way up to the order of a few thousand. As a result, unfortunately, it is not possible to compare the previous work and their achieved accuracy with each other objectively.

\section{\textbf{Methodology}} 

\subsection{\textbf{Dataset}}
Unlike DRDr, DRDR II did not enforce any mask requirement as a limitation for our dataset selection. And since we intended to use our pre-trained DRDr model as our primary feature extractor to generate the masks ourselves, we had the freedom to choose from all the huge datasets available out there for diabetic retinopathy that did not provide masks as a part of their data. To that end, we selected a public Kaggle dataset \cite{kdataset} that has more than 35 thousands fundus images with the size of 1024x1024. It was large enough for our task to ensure that overfitting would not occur and that our result could be reproducible \cite{farzan2019}. The dataset architecture was simple, it only provided the original fundus images, the eye that the picture was taken from (left vs right) along with a single integer indicating the severity of the case.

\subsection{\textbf{Preprocessing I}}
\label{p:prep1}
This new dataset, also, suffered from the same issues that were present in e-ophta \cite{decenciere2013teleophta}; the dataset which we had used for training DRDr. The images had been collected from different devices located in various photography facilities around the world and as a result, they had different pixel intensity values, lightings, and zoom levels. To counteract this issue, we employed the same preprocessing steps that we had used in DRDr as the starting point for our new preprocessing pipeline. For short, we employed OpenCV \cite{opencv_library} to crop the images and remove the extra blank space around them. Then, we transformed the eyes into perfect circles and removed the extra margins once more. To normalize the contrast level, we applied Gaussian blur (sigma X = 20) on each image and added the output to the original version. We assigned weights of 4 and -4 to the original and blurred images respectively. The gamma was also set to 128.

\subsection{\textbf{Data Frame Creation}}
After preparing the images it was time to bring our DRDr model into play. We decided to use our pre-trained model as the main feature extractor for the task at hand. The original DRDr was a Mask RCNN model that we had trained on the e-optha EX and e-optha MA datasets that contained images for exudates and microaneurysms or small hemorrhages respectively. However, we ended up having to change many hyper-parameters of the original implementation to make it usable for our task. An overview of the updated model hyper-parameters can be found in table \ref{table:Table1}.

\begin{table}
\centering
\caption{Hyper-parameters for Mask RCNN used for DRDr}
\resizebox{\columnwidth}{!}{%
\begin{tabular}{|>{\hspace{0pt}}m{0.545\linewidth}|>{\hspace{0pt}}m{0.39\linewidth}|} 
\toprule
\multicolumn{2}{|>{\centering\arraybackslash\hspace{0pt}}m{0.935\linewidth}|}{MASK RCNN} \\ 
\hline
Attribute & Value \\ 
\hline
BACKBONE & resnet101 \\ 
\hline
BACKBONE\_STRIDES & {[}4, 8, 16, 32, 64] \\ 
\hline
DETECTION\_MAX\_INSTANCES & 256 \\ 
\hline
DETECTION\_MIN\_CONFIDENCE & 0.35 \\ 
\hline
IMAGE\_MAX\_DIM & 1024 \\ 
\hline
IMAGE\_RESIZE\_MODE & square \\ 
\hline
LEARNING\_MOMENTUM & 0.9 \\ 
\hline
LEARNING\_RATE (Value, (\#Epochs)) & 10\textsuperscript{-4} (25), 10\textsuperscript{-5} (25), 10\textsuperscript{-6} (15) \\ 
\hline
MASK\_SHAPE & {[}28, 28] \\ 
\hline
MAX\_GT\_INSTANCES & 100 \\ 
\hline
NUM\_CLASSES & 3 (BG, EX, MA) \\ 
\hline
RPN\_ANCHOR\_RATIOS & {[}0.5, 1, 2] \\ 
\hline
RPN\_ANCHOR\_SCALES & (8, 16, 32, 64, 128) \\ 
\hline
RPN\_ANCHOR\_STRIDE & 1 \\ 
\hline
RPN\_TRAIN\_ANCHORS\_PER\_IMAGE & 512 \\ 
\hline
STEPS\_PER\_EPOCH & 500 \\ 
\hline
TOP\_DOWN\_PYRAMID\_SIZE & 256 \\ 
\hline
TRAIN\_BN & FALSE \\ 
\hline
TRAIN\_ROIS\_PER\_IMAGE & 512 \\ 
\hline
USE\_MINI\_MASK & FALSE \\
\bottomrule
\end{tabular}
}
\label{table:Table1}
\end{table}

Hence we geared up the model and initialized it with our pre-trained weights. Then we pumped all the images in the new dataset into it to get the masks for each instance of microaneurysm and exudate along with their confidence scores and bounding boxes. We used this info to conduct a data frame with the following details about each instance of the lesions found in an image: name of the file as the id, the eye that the image was taken from, the bounding box, its center position, the area of the masked lesion, type of the lesion, severity, and its confidence score. That left us with about 445 thousand instances collectively found for a total of 35 thousand images. 

It is also worth mentioning that DRDr normally takes a fundus image and the original mask as the input. But, at the test/production phase, the goal of the latter would only be to help the model illustrate the differences between the predicted mask and the original one. Since this fancy feature is not needed for our new task at hand, we used a fixed black binary mask and passed it alongside all of the images in our dataset.

\subsection{\textbf{Preprocessing II}}
In our preceding paper \cite{shenavarmasouleh2020drdr}, we have elaborated on why we had to tweak the confidence score to get the most out of our model. DRDr had originally been trained to output all the instances that it is more than 35\% confident about. It worked well for our previous task, however, in DRDr II, we realized that it is not good enough and it can add unwanted noise to the dataset. This was mainly an issue for the exudate instances since their areas were typically several folds larger than microaneurysms. Hence, as the first step towards putting the data frame created in the previous phase into use, we pruned all the exudate instances with a confidence score of below 65\%. Next, we converted the eye column (left vs right) into categorical entities and then calculated weighted areas (confidence score multiplied by area) to counteract the big areas with low probabilities.

Originally, the images in the dataset were separated into five categories. But, based on the case study conducted by Google \cite{tfstudy}, the are so many overlaps in the adjacent categories in terms of severity that even experts were not sure which category to pick as the final choice. Hence, we grouped the top two most and bottom two most groups and reduced the number of classes to three to make up for it.

Next, to bundle all the various instances of lesions found in one image together to form one single row in the data frame for each image found in the dataset, we grouped all the rows based on their name and then calculated the following attributes for each of them: number of instances per lesion type (EX and MA), the sum of weighted area for each type, and mean and standard deviation for the center of bounding boxes for each lesion type. To further improve our model and make the system robust to the outliers we then proceeded and calculated the z-score for the aforementioned columns and dropped the instances that were not within two standard deviations.

We started using the data frame at that point, but then we realized that the dataset was not well balanced as the majority of images did not have any issue (the patient was healthy) and that caused the model to get a very good accuracy just by outputting 0 all the time without learning any useful relationship among the features. Hence, we decided to under-sample the 0 class to make the final dataset well balanced. 

We then normalized all the attributes in order to make them be in the range of 0 to 1 that is proven to help the model to converge faster. At this point, the data frame was ready and each row in it corresponded to an image in the original dataset. We then shuffled the whole thing and since the data frame was large enough, we decided to use 80\% of the rows for the training phase and the remainder for validation and testing phases equally.

\section{\textbf{Experiments and Results}}
As for our main classifier, we used a feed-forward neural network with 2 hidden states each with 75 nodes followed by a softmax layer that outputted the corresponding severity level. We used Adam optimizer with a learning rate of 0.001 and trained the model for 100 epochs. To compare it with other classifiers and get a sense of where we stand among other machine learning approaches mentioned in the related works, we also trained Linear SVM, Kernel SVM (polynomial and RBF), Logistic Regression, Decision Tree, AdaBoost, Naïve Bayes, and KNN. You can find our results in table \ref{table:Table2}. Unlike most of the related work, we did not ensemble any of our classifiers together. However, for the most part, the margins of differences were significantly narrow and consequently, it is unlikely that ensembling could lead to any significant positive difference.

We also performed a brief ablation study over our data frame to find out the amount of influence each feature has over the final accuracy score. According to our findings, all columns played a positive role and had a positive correlation with the final result, however, the most dominant one proved to be the sum of weighted areas for each type of the lesions.

This was also a great test for our DRDr model since the performance of segmentation models are often measured with Intersection over Union (IoU) and mean Average Precision (mAP), and although they are both great measures, they are not as intuitive as the traditional accuracy score. It can be perceived that DRDr proved itself once again and showed that in addition to being able to find lesions in the fundus images, it can also be used as a solid and reliable feature extractor to help find the severity of infamous diabetic retinopathy.

\begin{table}
\centering
\caption{DRDr II results for different classifiers}
\begin{tabular}{|>{\hspace{0pt}}m{0.58\linewidth}|>{\hspace{0pt}}m{0.29\linewidth}|} 
\toprule
Classifier & Accuracy (\%) \\ 
\hline
Neural Network & 92.55 \\ 
\hline
Decision Tree & 91.17 \\ 
\hline
Naïve Bayes & 85.25 \\
\hline
SVM - Polynomial & 79.10 \\ 
\hline
Logistic Regression & 78.99 \\ 
\hline
SVM - Linear & 78.65 \\ 
\hline
SVM - RBF & 77.20 \\ 
\hline
KNN & 75.86 \\ 
\hline
AdaBoost & 60.00 \\ 
\bottomrule
\end{tabular}
\label{table:Table2}
\end{table}

\section{\textbf{Conclusion and Future Work}}
In this paper, we presented DRDr II, a proceeding work built on top of DRDr, to aid doctors and clinicians in the diagnosis of Diabetic Retinopathy and help them identify the severity of the cases. We hypothesize that the accuracy can further be improved by also using CNN features extracted from the images or masks and fusing them directly with our current features \cite{Asali2020DeepMSRFAN}. Also, contextual info can play a significant role in the diagnosis as well, but unfortunately, not many datasets provide such information. In future work, we plan to address the forenamed issues and we will also try to unify DRDr and DRDr II and morph them into a single monolithic deep learning model so that it can produce binary masks for the lesions and find the severity level at the same time.


\section{\textbf{Conflict of Interest}}
The authors declare that there is no conflict of interest regarding the publication of this article.


\bibliography{bib} 
\bibliographystyle{ieeetr}

\end{document}